\documentclass{pasa}%
\usepackage{lscape, morefloats, graphicx}
\title{Bridging the Ultraviolet and Optical Regions: Transformation Equations between {\it GALEX} and {\it UBV} Photometric Systems}
\author[Bilir et al.]{S. Bilir$^1$\thanks{sbilir@istanbul.edu.tr}, N. Alan$^1$, S. Tun\c cel G\"u\c ctekin$^1$, M. \c Celebi$^2$, T. Yontan$^2$, O. Plevne$^2$, S. Ak$^1$, T. Ak$^1$, S. Karaali$^1$\\
     \affil{$^1$Istanbul University, Faculty of Science, Department of Astronomy and Space Sciences, 34119, Beyaz\i t, Istanbul, Turkey}
     \affil{$^2$Istanbul University, Institute of Graduate Studies in Science, Programme of Astronomy and Space Sciences, 34116, Beyaz{\i}t, Istanbul, Turkey\\}
}
\jid{PASA}
\doi{10.1017/pas.\the\year.xxx}
\jyear{\the\year}
\usepackage{natbib}
\usepackage{graphicx}
\begin{document}
\begin{abstract}
We derive transformation equations between {\it GALEX} and {\it UBV} colours by using the reliable data of 556 stars. We present two sets of equations; as a function of (only) luminosity class, and as a function of both luminosity class and metallicity. The metallicities are provided from the literature, while the luminosity classes are determined by using the PARSEC mass tracks in this study. Small colour residuals and high squared correlation coefficients promise accurate derived colours. The application of the transformation equations to 70 stars with reliable data shows that the metallicity plays an important role in estimation of more accurate colours.      
\end{abstract}
\begin{keywords}
techniques: photometric - catalogue - surveys 	
\end{keywords}
\maketitle%

\section{Introduction}
Reliable spectroscopic, photometric and astrometric data are important for understanding the structure, formation and evolution of our Galaxy. Today, sky surveys are systematically carried out in a wide range of the electromagnetic spectrum from X-rays to the radio. In some sky surveys such as {\it ROSAT} \citep{Snowden95}, {\it GALEX} \citep{Martin05}, and SDSS \citep{York00} which are carried out between the X-ray and optical regions of the electromagnetic spectrum, the interstellar absorption prevents to obtain accurate magnitude and colours. However, this problem has been overcame by means of the sky surveys which are defined over the infrared region of the star spectrum, i.e. 2MASS \citep{Skrutskie06}, UKIDSS \citep{Hewett96}, VVV \citep{Minniti10}, VISTA \citep{Cross12}, {\it WISE} \citep{Wright10}, and {\it AKARI} \citep{Murakami07}. Thus, considerable information have been obtained, especially on the bulge, bar structure and Galactic stellar warp in our Galaxy \citep{Dwek95, Lopez-Corredoira05, Lopez-Corredoira19a, Lopez-Corredoira19b, Benjamin05}.

Today, photometric sky survey observations are systematically performed to cover the ultraviolet (UV) and infrared regions of the electromagnetic spectrum. In spectroscopic sky surveys, spectroscopic observations are made for a limited number of objects with different luminosities which are classified according to their positions in colour spaces obtained from the photometric observations. The main-sequence stars provide information about the solar neighbourhood and the evolved stars provide information about the old thin disc, thick disc and halo populations of our Galaxy beyond the solar vicinity. The number of stars observed on current spectroscopic sky survey programs do not exceed one million in total. As this number is less than needed for a detailed study of the Galactic structure, precise measurements of photometric sky surveys, which contain billions of bright and faint objects, are still important in testing models about the structure and evolution of the Galaxy. 

\begin{figure*}
\centering
\includegraphics[width=16cm,height=6cm, keepaspectratio]{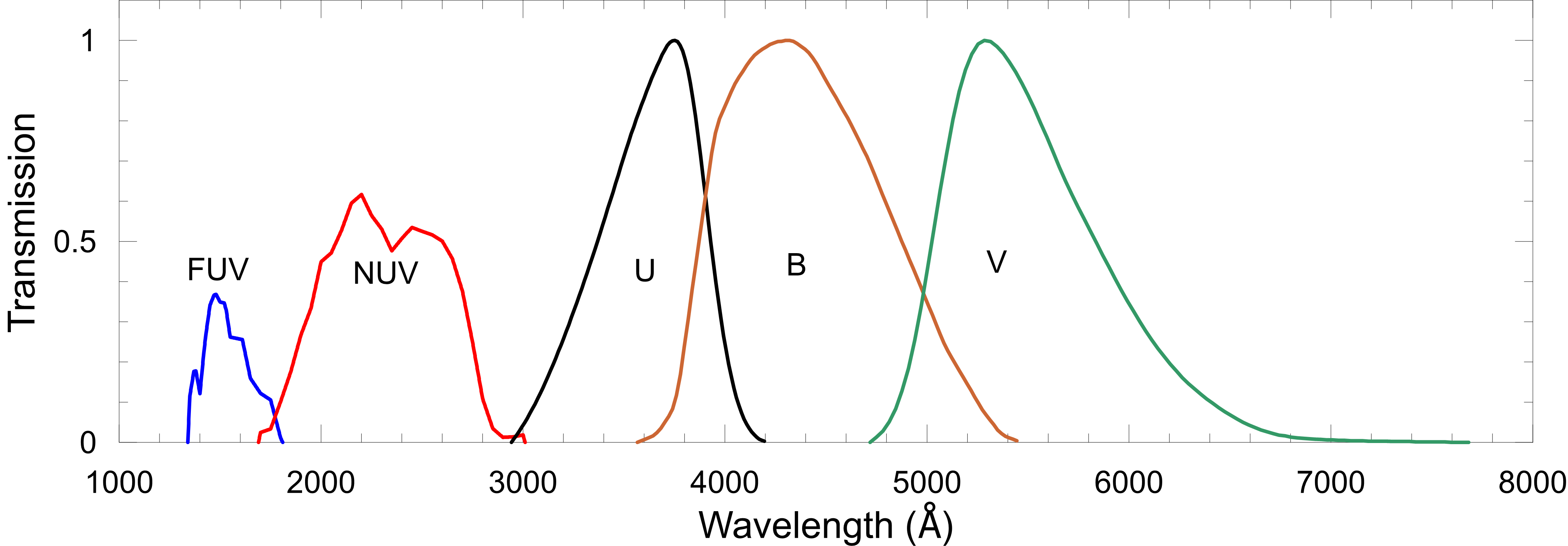}
\caption{Normalized transmission curves of the {\it GALEX} $FUV$, $NUV$ and Johnson-Morgan $U$, $B$, $V$ filters.}
\end{figure*} 

Photometric sky surveys performed in different regions of the electromagnetic spectrum are designed to include shallow or deep magnitudes, according to the purpose of the researchers. However, transformations between different photometric systems can be used as a tool to combine shallow and deep magnitudes. These transformation equations can also be produced in terms of luminosity and metallicity. In the literature, transformation equations are given for main-sequence stars \citep{Smith02, Karaali05, Bilir05, Bilir08a, Bilir11, Rodgers06, Jordi06, Covey07, Chonis08} and evolved stars \citep{Straizys09, Yaz10, Bilir12, Bilir13, Ak14}. Transformation equations between the photometric systems have been obtained for approximately 20 years and used effectively to investigate the structure and evolution of our Galaxy. {\it UBV} is one of the important photometric system in the optical region \citep{Johnson53} which is used to determine the photometric metallicities of the stars and the interstellar absorption. The $U-B$ colour index plays an important role in these determinations, while its combination with the colour index $B-V$ can be used in determination of the reddening of stars. In the {\it UBV} photometric system, the colour excess $E(B-V)$ of a star can be determined by the $Q$ method or by shifting its observed $B-V$ colour index along the reddening vector in the $U-B\times B-V$ two-colour diagram up to the intrinsic colour index $(B-V)_0$ \citep{Johnson53}. While the colour excess $E(U-B)$ can be calculated by the equation of the reddening line, i.e. $E(U-B)=0.72\times E(B-V)+0.05\times E(B-V)^2$. \citet{Roman55} discovered that stars with weak metal lines have larger ultraviolet (UV) excesses than the ones with strong metal lines. \citet{Schwarzschild55}, \citet{Sandage59}, and \citet{Wallerstein62} confirmed the work of \citet{Roman55}. Thus, metal-rich and metal-poor stars could be classified not only spectroscopically, but also photometrically, i.e. by their UV excesses. \citet{Sandage69} noticed in a $(U-B)_0\times(B-V)_0$ two-colour diagram of a set of stars in solar neighbourhood that stars with the same metallicity have a maximum UV excess at the colour index $(B-V)_0=0.6$ mag, and introduced a procedure to reduce the UV excesses of the stars to the one of colour index $(B-V)_0=0.6$ mag. The relation between the UV excess and metallicities of stars have been used by the researchers for their metallicity estimation via photometry i.e. \citep{Carney79, Karaali03a, Karaali03b, Karaali03c, Karatas06, Karaali11, Tuncel16, Celebi19}. Similar calibrations have been developed for the SDSS photometric system and applied to faint stars \citep{Karaali05, Bilir05, Tuncel17}. These calibrations were used to calculate metal abundances \citep{Ak07, Ivezic08, Tuncel19} and estimation of the Galactic model parameters for different populations \citep{Karaali04, Ak07, Bilir08b, Juric08, Yaz-Karaali10}.

The {\it UBV} photometric system provides reliable data for the bright stars which occupy the solar neighbourhood. However, for the faint stars the same case thus not hold. Additionally, the low transmission of the Earth's atmosphere limits the number of stars with reliable $U$ magnitudes. This problem could be solved by measurements performed outside of the atmosphere such as the satellite Galaxy Evolution Discovery ({\it GALEX}) \citep{Martin05}.

The {\it GALEX} satellite was launched in 2003 and continued its active mission until 2012. It is the first satellite to observe the entire sky with the two detectors, i.e. far ultraviolet ($FUV$, $\lambda_{eff}$ = 1528\AA; 1344 - 1786\AA) and near ultraviolet ($NUV$, $\lambda_{eff}$ = 2310\AA; 1771 - 2831\AA). The passbands of {\it GALEX} and {\it UBV} photometric systems are shown in Fig. 1. Measurements in the far and near UV bands of approximately 583 million objects obtained from the reduction of 100865 images from satellite observations are given in DR 6+7 versions of {\it GALEX} database \citep{Bianchi17}. In our study, transformation equations between the colour indices of {\it GALEX} and {\it UBV} photometric systems are derived in terms of the luminosity class. These equations provide us empirical $U-V$ colour index and UV excess for stars which can be used in photometric metal abundance estimation. We organized the paper as follows. Section 2 is devoted to the selection of the calibration stars in our study, derivation of the calibrations is given in Section 3, and finally the results and discussion are presented in Section 4.

\begin{figure}[t]
\centering
\includegraphics[width=6.5cm,height=8.67cm,]{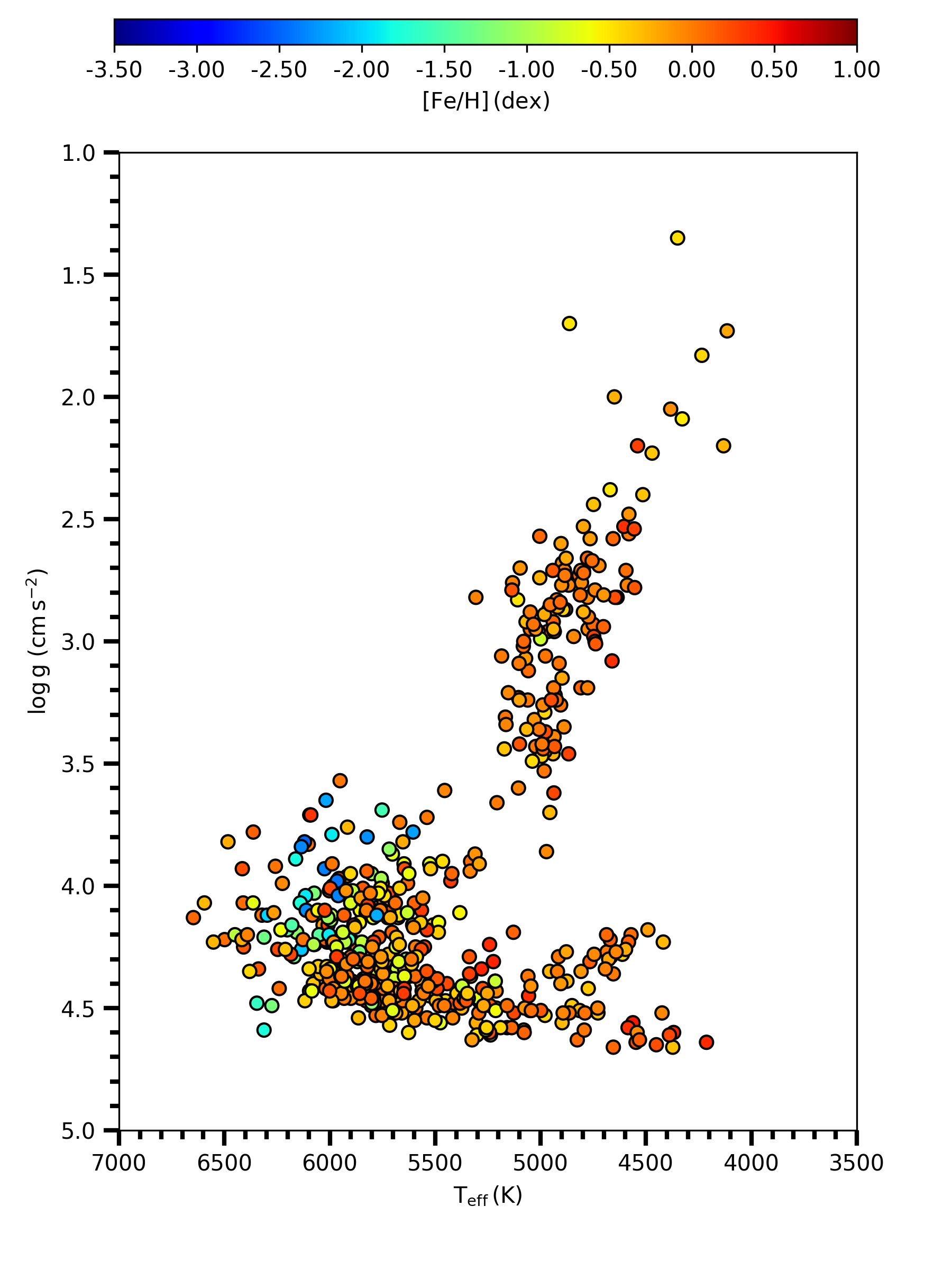}
\caption{$\log g \times T_{eff}$ diagram of the stars in the sample. The diagram is colour-coded for the metallicity of 556 stars.}
\end{figure} 

\section{Data}
In this study, we prioritized the stars that have precise spectroscopic, astrometric and photometric data in the literature. In this context, we used the spectroscopic data from 14 studies \citep{Boesgaard11, Nissen11, Ishigaki12, Mishenina13, Molenda13, Bensby14, daSilva15, Sitnova15, Jofre15, Brewer16, Kim16, Maldonado16, Luck17, DelgadoMena17}. 6149 stars with atmospheric model parameters ($T_{eff}$, $\log g$ and [Fe/H]) could be provided from these studies (Table 1). The photometric data are supplied from {\it GALEX} DR7 \citep{Bianchi17} and {\it UBV} \citep{Oja84, Mermilliod87, Mermilliod97, Ducati02, Koen10, Carrasco10}, while the trigonometric parallaxes are taken from {\it Gaia} DR2 \citep{Gaia18}. The photometric and astrometric data of 5593 stars were not available for the original set of stars (6149 stars). Hence, our sample reduced to 556 (Table 2). 

The $\log g \times T_{eff}$ diagram of the sample stars is shown in Fig. 2 with the colour coded for metallicity [Fe/H]. PARSEC mass tracks for different metal abundances are used to determine the luminosity classes of the sample stars \citep{Bressan12, Tang14, Chen15}. The evolutionary tracks generated for different heavy element abundances ($Z$ = 0.040, 0.030, 0.020, 0.017, 0.014, 0.010, 0.008, 0.006, 0.004 and 0.002) are converted to [Fe/H] metallicities using the formulae given by Jo Bovy\footnote{https://github.com/jobovy/isodist/blob/master/isodist/\\Isochrone.py} \citep[see also,][]{Bostanci18, Eker18, Yontan19, Banks20}. Then, zero age main sequence (ZAMS) and terminal age main sequence (TAMS) evolutionary tracks  corresponding to the metal abundance ranges for mean $Z$ values were established (Fig. 3). The luminosity classes of the sample stars were determined by the metallicity intervals as indicated in Fig. 3 and marked on the $\log g \times T_{eff}$ diagrams. Stars between the ZAMS and TAMS curves are classified as main-sequence stars, while the ones above the TAMS curve are adopted as evolved stars, i.e. those with $\log g\geq 3.5$ as sub-giants and the ones with $\log g<3.5$ as giants. Thus, the number of stars with different luminosity classes turned out to be as follows: 245 main-sequence, 187 sub-giants, and 124 giants. 

We used the atmospheric model parameters to classify the luminosity class of each star. The giant stars tend to be well separated from the sub-giants, while the sub-giants are very close to the main-sequence stars. Hence, we investigated the uncertainty of the atmospheric model parameters to reveal any contamination of the sub-giants into the main sequence region and vice-versa, as explained in the following. As the uncertainty of the atmospheric model parameters were not considered in some studies which cover our sample stars (556 stars), we used the uncertainty of the atmospheric model parameters in \citet{Bensby14} which contains approximately 22\% of the stars in the sample, for our purpose. The median errors of $T_{eff}$, $\log g$ and [Fe/H] in \citet{Bensby14} are 56 K, 0.08 cm s$^{-2}$ and 0.05 dex, respectively. The luminosity classes of the stars in the sample were determined by comparing the atmospheric model parameters of the stars with the ZAMS and TAMS curves designated from the PARSEC mass tracks. The median errors of the atmospheric model parameters were added to the original parameters of the stars their luminosity classes were re-asigned. Stars, whose luminosity classes were changed, were considered as contamination. It is found that, the contamination of the main-sequence stars by sub-giant stars is 2.9\%, the contamination of sub-giant stars by main-sequence is 3.7\%, and the contamination of giant stars by sub-giant stars is 3.2\%. Hence, one can say that our transformation equations can be considered for the luminosity class of the sample stars in question. 

\begin{table*}[htbp]
\setlength{\tabcolsep}{2pt}
{\tiny
  \caption{Spectroscopic data used in this study. $N$ denotes the number of stars, $R$ spectral resolution, $S/N$ signal-to-noise ratio. Observatory, telescope and the spectrograph used in the observations are also noted.}
    \begin{tabular}{clcccl}
\hline
    ID & Authors & $N$ & $R$ & $S/N$ & Observatory / Telescope / Spectrograph \\
\hline
 1 & \citet{Boesgaard11} & 117 & $\sim$42000 &   106  & Keck / Keck I / HIRES \\
 2 & \citet{Nissen11}    & 100 & 55000       & 250-500& ESO / VLT / UVES, ORM / NOT / FIES \\
 3 & \citet{Ishigaki12}  &  97 & 100000      & 140-390& NAOJ / Subaru / HDS \\
 4 & \citet{Mishenina13} & 276 & 42000       &  $>100$& Haute-Provence / 1.93m / ELODIE \\
 5 & \citet{Molenda13}   & 221 & 25000-46000 & 80-6500& ORM / NOT / FIES, OACt / 91cm / FRESCO, ORM / Mercator / HERMES\\
   &  &  &  &  & OPM / TBL / NARVAL, MKO / CFHT / ESPaDOnS \\
 6 & \citet{Bensby14}    & 714 & 40000-110000& 150-300& ESO / 1.5m and 2.2m / FEROS, ORM / NOT / SOFIN and FIES, \\
   &  &  &  &  &                                     ESO / VLT / UVES, ESO / 3.6m / HARPS, Magellan Clay / MIKE \\
 7 & \citet{daSilva15}   & 309 & $\sim$42000 & $>150$ & Haute Provence / 1.93m / ELODIE \\
 8 & \citet{Sitnova15}   &  51 & $>60000$    & 70-100 & Lick / Shane 3m / Hamilton, CFH / CFHT / ESPaDOnS \\
 9 & \citet{Jofre15}     & 223 &$30000-120000$& $>150$ & ESO / 3.6m / HARPS, ESO / 2.2m / FEROS, OHP / 1.93m / ELODIE,\\ 
   &  &  &  &  & OHP / 1.93m / SOPHIE, CASLEO, 2.15m / EBASIM\\
10 & \citet{Brewer16}    &1615 & $\sim$70000 & $>200$ & Keck / Keck I / HIRES \\
11 & \citet{Kim16}       &170  & 10000       & $>100$ & KPNO / Mayall 4m / Echelle spectrograph\\
12 & \citet{Maldonado16} & 154 & $\sim$42000-115000 & 107 & La Palma / Mercator / HERMES, ORM / NOT / FIES, \\
   &  &  &  &  & Calar Alto / 2.2m / FOCES, ORM / Nazionale Galileo / SARG \\
13 & \citet{Luck17}      &1041 & 30000-42000 & $>75$  & McDonald / 2.1m / SCES, McDonald / HET / High-Resolution \\
14 &\citet{DelgadoMena17}&1059 & $\sim$115000 & $>200$ & HARPS GTO programs \\
\hline
    \end{tabular}%
\\
CASLEO: Complejo Astronomico El Leoncito, EBASIM: Echellede Banco Simmons, CFH: Canada-France-Hawaii, CFHT: Canada-France-Hawaii Telescope, 
ESO: European Southern Observatory, ESPaDOnS: an Echelle SpectroPolarimetric Device for the Observation of Stars at CFHT, 
FEROS: The Fiber-fed Extended Range Optical Spectrograph, FIES: The high-resolution Fibre-fed Echelle Spectrograph,
FOCES: a fibre optics Cassegrain echelle spectrograph, FRESCO: Fiber-optic Reosc Echelle Spectrograph of Catania Observatory, 
GTO: Guaranteed Time Observations, HARPS: High Accuracy Radial velocity Planet Searcher, HERMES: High-Efficiency and high-Resolution Mercator Echelle Spectrograph, 
HET: Hobby-Eberly Telescope, HDS: High Dispersion Spectrograph, HIRES: High Resolution Echelle Spectrometer, KPNO: Kitt Peak National Observatory,
MIKE: Magellan Inamori Kyocera Echelle, MKO: Mauna Kea Observatory, NAOJ: National Astronomical Observatory of Japan, NOT: Nordic Optical Telescope, 
OACt: Catania Astrophysical Observatory, OPM: Observatorie Pic du Midi, ORM: Observatorio del Roque de losMuchachos, SCES: Sandiford Cassegrain Echelle Spectrograph,
SOFIN: The Soviet-Finnish optical high-resolution spectrograph, SOPHIE: Spectrographe pour l'Observation des Phenomenes des Interieurs stellaires et des Exoplanetes, 
TBL: Telescope Bernard Lyot, UVES: Ultraviolet and Visual Echelle Spectrograph, VLT: Very Large Telescope, 
\label{tab:addlabel}%
}
\end{table*}%

\begin{table*}[htbp]
  \centering
\setlength{\tabcolsep}{1.7pt}
{\scriptsize
  \caption{The basic parameters of 556 sample stars; ID, star, equatorial coordinates in J2000 ($\alpha$, $\delta$), photometric data ($FUV$, $NUV$, $V$, $U-B$, $B-V$), reduced colour excess ($E_d(B-V)$), atmospheric model parameters ($T_{eff}$, $\log g$ and [Fe/H]) and their references, and trigonometric parallaxes ($\pi$) with the errors taken from {\it Gaia} DR2.}
    \begin{tabular}{clcccccccccccccc}
\hline   
    ID & Star & $\alpha$ & $\delta$& $FUV$ & $NUV$ & $V$ & $U-B$ & $B-V$ & $E_d(B-V)$ & $T_{eff}$ & $\log g$ & [Fe/H] & Reference & $\pi$ & $\sigma_{\pi}$ \\
      &  & (hh:mm:ss) & (dd:mm:ss) & (mag) & (mag) & (mag) & (mag) & (mag) & (mag) & (K) & (cm s$^{-2}$) & (dex) &   & (mas) & (mas) \\
\hline
    1 & Hip 80 & 00 00 58.28 & $-$11 49 25.50 & 20.026 & 13.123 & 8.400 & -0.080 & 0.550 & 0.012 & 5856 & 4.10 & -0.59 & (6) & 13.9286 & 0.0691 \\
    2 & HD 225197 & 00 04 19.79 & $-$16 31 44.50 & 21.159 & 14.675 & 5.780 & 1.054 & 1.080 & 0.012 & 4778 & 2.66 & 0.11 & (7) & 9.8054 & 0.0946 \\
    3 & HD 249 & 00 07 22.56 & +26 27 02.20 & 22.101 & 15.353 & 7.381 & 0.750 & 0.995 & 0.015 & 4775 & 2.95 & -0.04 & (10) & 7.6866 & 0.0371 \\
    4 & HD 870 & 00 12 50.25 & $-$57 54 45.40 & 19.946 & 13.660 & 7.226 & 0.344 & 0.774 & 0.001 & 5381 & 4.42 & -0.10 & (14) & 48.4741 & 0.0263 \\
    5 & Hip 1128 & 00 14 04.48 & $-$11 18 41.70 & 21.027 & 13.565 & 8.360 & 0.015 & 0.655 & 0.009 & 5522 & 4.37 & -0.64 & (6) & 23.3418 & 0.0534 \\
    ... & ... & ... & ... & ... & ... & ... & ... & ... & ... & ... & ... & ... & ... & ... & ... \\
    ... & ... & ... & ... & ... & ... & ... & ... & ... & ... & ... & ... & ... & ... & ... & ... \\
    ... & ... & ... & ... & ... & ... & ... & ... & ... & ... & ... & ... & ... & ... & ... & ... \\
    552 & Hip 117526 & 23 50 05.74 & +02 52 37.82 & 21.482 & 14.599 & 8.339 & 0.362 & 0.744 & 0.010 & 5540 & 4.35 & 0.20 & (6) & 21.3273 & 0.0588 \\
    553 & HD 223524 & 23 50 14.73 & $-$09 58 26.90 & 21.009 & 14.832 & 5.941 & 1.150 & 1.130 & 0.015 & 4656 & 2.58 & 0.10 & (13) & 10.4176 & 0.0925 \\
    554 & Hip 117946 & 23 55 26.60 & +22 11 35.80 & 20.229 & 16.266 & 8.770 & 0.810 & 1.020 & 0.006 & 4790 & 4.52 & 0.04 & (10) & 39.1780 & 0.0580 \\
    555 & Hip 118115 & 23 57 33.52 & $-$09 38 51.10 & 20.277 & 13.434 & 7.863 & 0.146 & 0.641 & 0.011 & 5833 & 4.39 & 0.02 & (6) & 19.4670 & 0.0642 \\
    556 & Hip 118278 & 23 59 28.43 & $-$20 02 05.00 & 20.803 & 14.054 & 7.470 & 0.290 & 0.740 & 0.004 & 5533 & 4.41 & -0.07 & (11) & 38.1312 & 0.0519 \\
\hline
    \end{tabular}%
  \label{tab:addlabel}%
}
\end{table*}%

\begin{landscape}
\textwidth = 750pt
\begin{figure*}
\centering
\includegraphics[width=24cm, height=12.8cm, keepaspectratio]{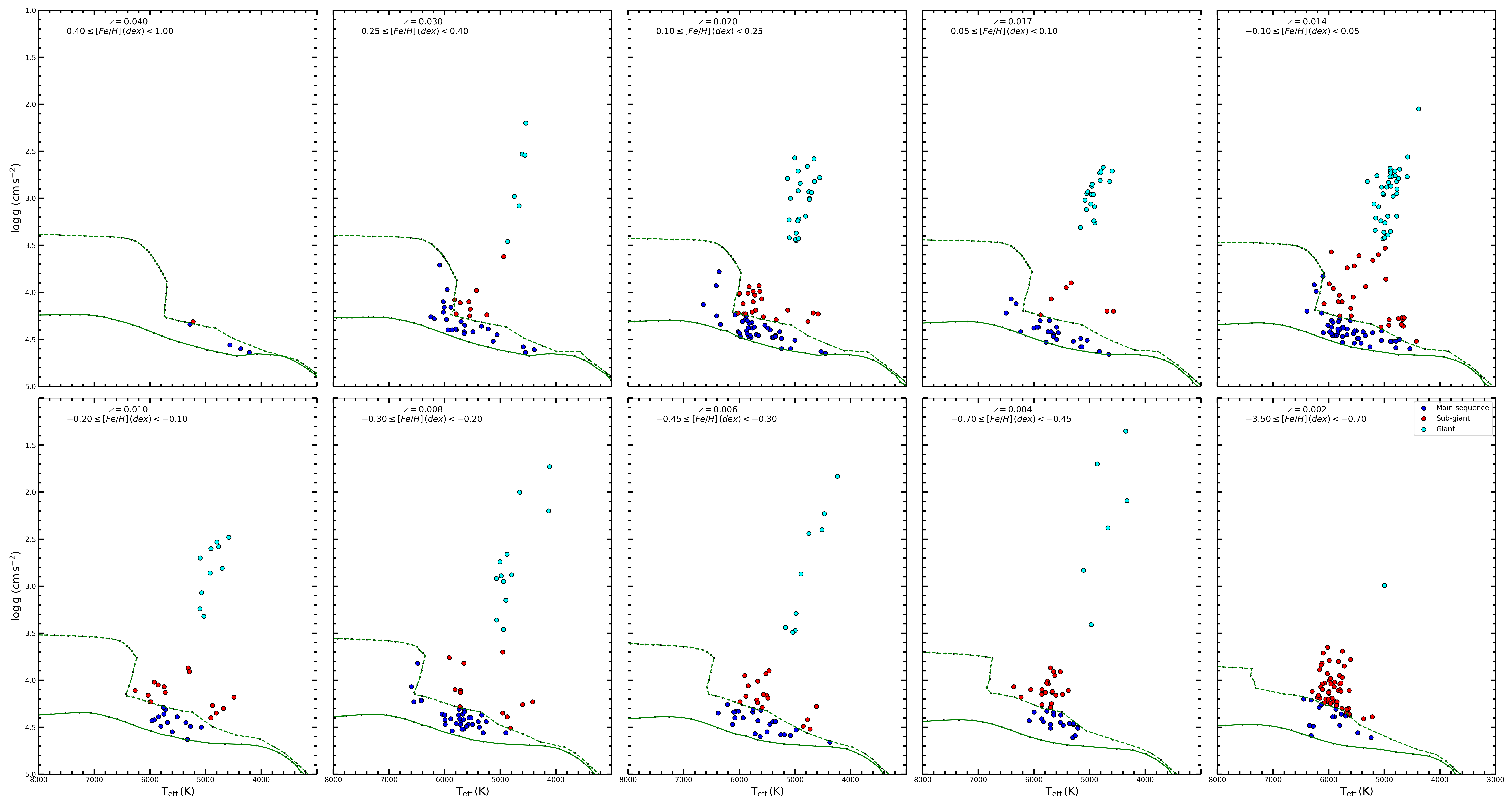}
\caption{$\log g \times T_{eff}$ diagram of the stars with different metallicity intervals. Blue circle: main-sequence, red circle: sub-giants and cyan circle: giant stars. Green solid and dashed curves represent the ZAMS and TAMS, respectively.}
\end{figure*} 
\end{landscape}

\begin{figure}[ht]
\centering
\includegraphics[width=5cm,height=10cm,keepaspectratio]{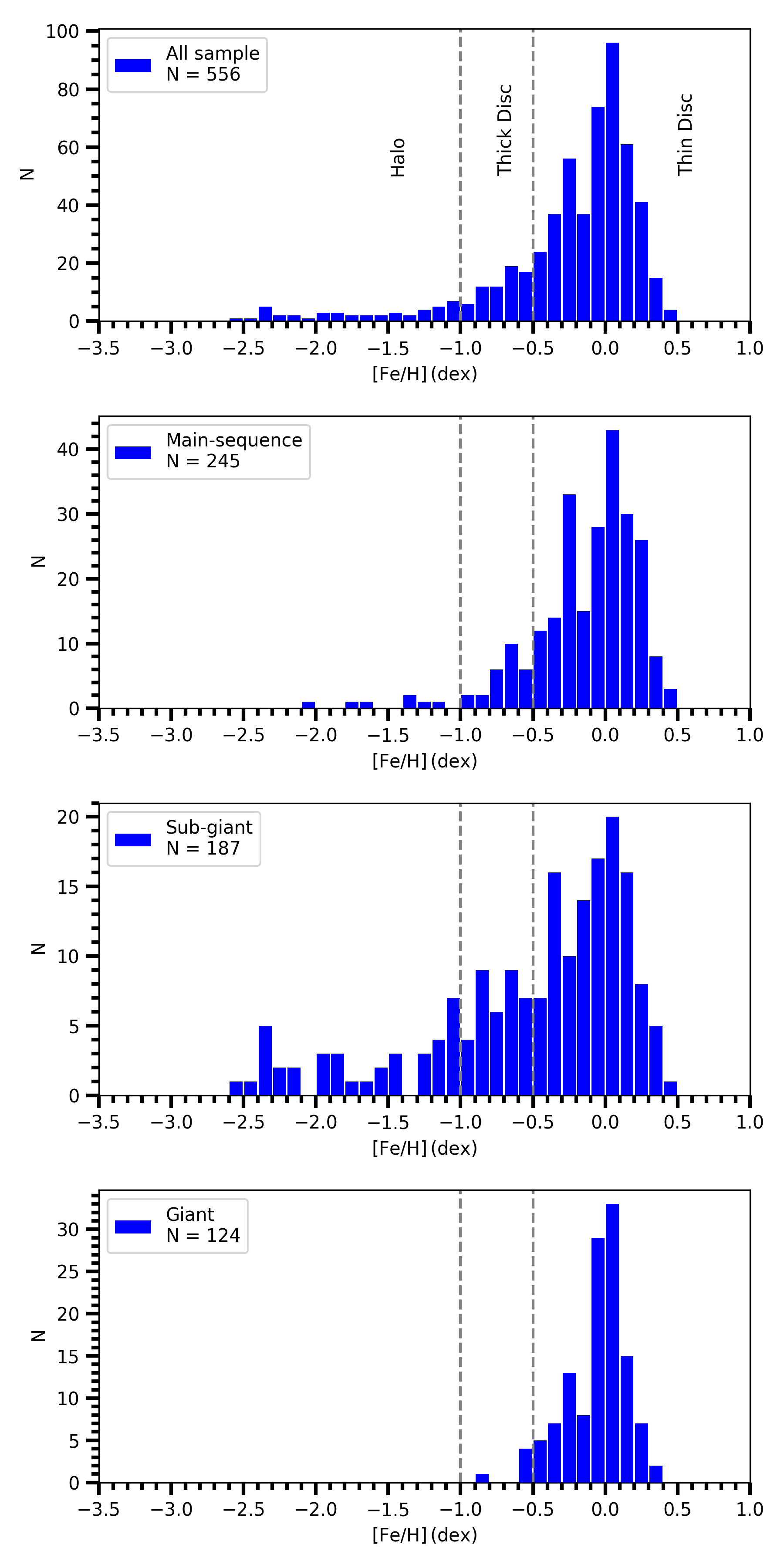}
\caption{Distribution of the spectroscopic metal abundances for all sample, main-sequence, sub-giant, and giant stars.}
\end{figure} 

\begin{figure}[hb]
\centering
\includegraphics[width=7cm,height=7cm,keepaspectratio]{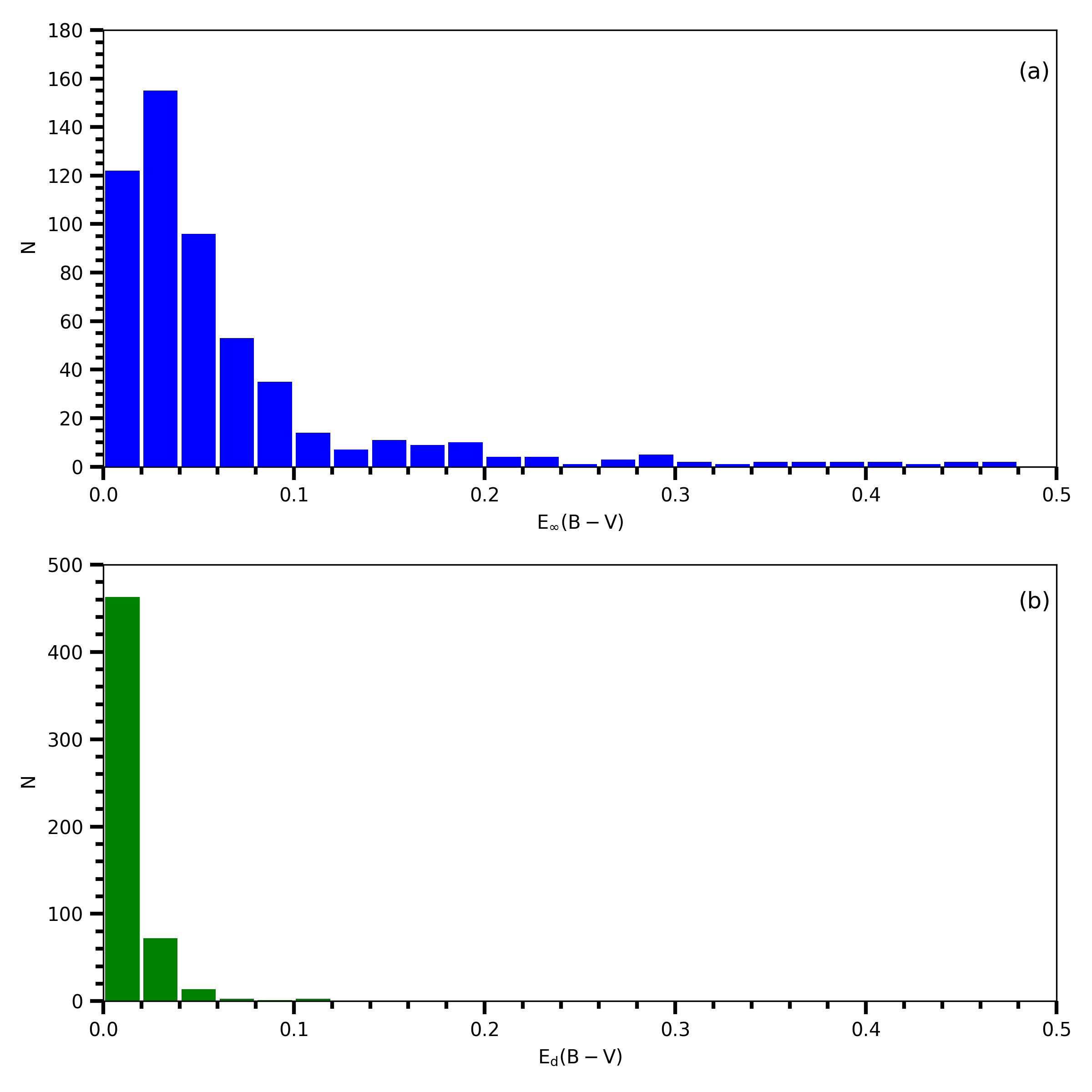}
\caption{Histograms of the original $E_{\infty}(B-V)$ (a) and reduced $E_d(B-V)$ (b) colour excesses of 556 stars.}
\end{figure} 

\begin{table*}[htbp]
  \centering
  \caption{Distribution of 556 sample stars according to the luminosity classes and the metallicity intervals.}
    \begin{tabular}{lcccc}
\hline
     & ${\rm [Fe/H]}\leq -1$ & $-1 <{\rm[Fe/H]}\leq -0.5$ & $-0.5<{\rm [Fe/H]}\leq +0.5$ & Total \\
     & (dex) & (dex) & (dex) &  \\
\hline
    Main-sequence & 7 & 26 & 212 & 245 \\
    Sub-giant & 38 & 35 & 114 & 187 \\
    Giant & -- & 5 & 119 & 124 \\
\hline
    \end{tabular}%
  \label{tab:addlabel}%
\end{table*}%

The sample stars were also separated into different population types according to their metallicities, i.e. thin disc ($-0.5 <{\rm[Fe/H]}<+0.5$ dex), thick disc ($-1<{\rm[Fe/H]}<-0.5$ dex), and halo (${\rm [Fe/H]}<-1$ dex), and they were listed in Table 3. The metallicity distribution for all stars and for different luminosity classes are given in Fig. 4. 

We estimated the interstellar absorption in the {\it UBV} system for the sample stars by using the dust map of \citet{Schlafly11} and reduced it to the distance of the star in question by the following equation of \citet{Bahcall80}:

\begin{equation}
A_{d}(b)=A_{\infty}(b)\Biggl[1-\exp\Biggl(\frac{-\mid d~\sin(b)\mid}{H}\Biggr)\Biggr].
\end{equation}   	
Here, $d$ and $b$ are the distance and Galactic latitude of a star, respectively, and $H$ indicates the scaleheight of the Galactic dust \citep[$H=125$ pc;][]{Marshall06}. Distances of the stars are calculated by using the trigonometric parallaxes in {\it Gaia} DR2 via the equation $d{\rm(pc)}=1000/\pi$(mas). The median distances of the main-sequence, sub-giant and giant stars are 33, 49 and 83 pc, respectively. Thus, the colour excesses $E_d(B-V)$ and $E_d(U-B)$ could be calculated by replacing the total absorption $A_d(b)$ in the following equations of \citet{Cardelli89} and \citet{Garcia88}:
{\small
\begin{eqnarray}
E_d(B-V)=A_d(b)/3.1\\ \nonumber
E_d(U-B)=0.72\times E_d(B-V)+0.05\times E_d(B-V)^2
\end{eqnarray}}
Then, we estimated the intrinsic colours and magnitudes by using the following equations. Extinction coefficients in the equations are taken from  \citet{Cardelli89} and \citet{Yuan13}:

\begin{eqnarray}
(B-V)_0=(B-V)-E_d(B-V)\\\nonumber
(U-B)_0=(U-B)-E_d(U-B)\\\nonumber
V_0=V-3.1\times E_d(B-V)\\\nonumber
(FUV)_0=FUV-4.37\times E_d(B-V)\\\nonumber
(NUV)_0=NUV-7.06\times E_d(B-V)\\\nonumber
\end{eqnarray}
Distribution of the colour excesses $E_{\infty}(B-V)$ and $E_d(B-V)$ are plotted in Fig. 5. Small colour excesses indicate that our star sample consist of solar neighbourhood stars. The two-colour diagrams, $(U-V)_0\times(B-V)_0$ and $(U-V)_0\times(FUV-NUV)_0$, of the sample stars are plotted in Fig. 6 and Fig. 7 with colour coded for the luminosity class and metallicity, respectively. 

\begin{figure*}
\centering
\includegraphics[width=12cm,height=12cm,]{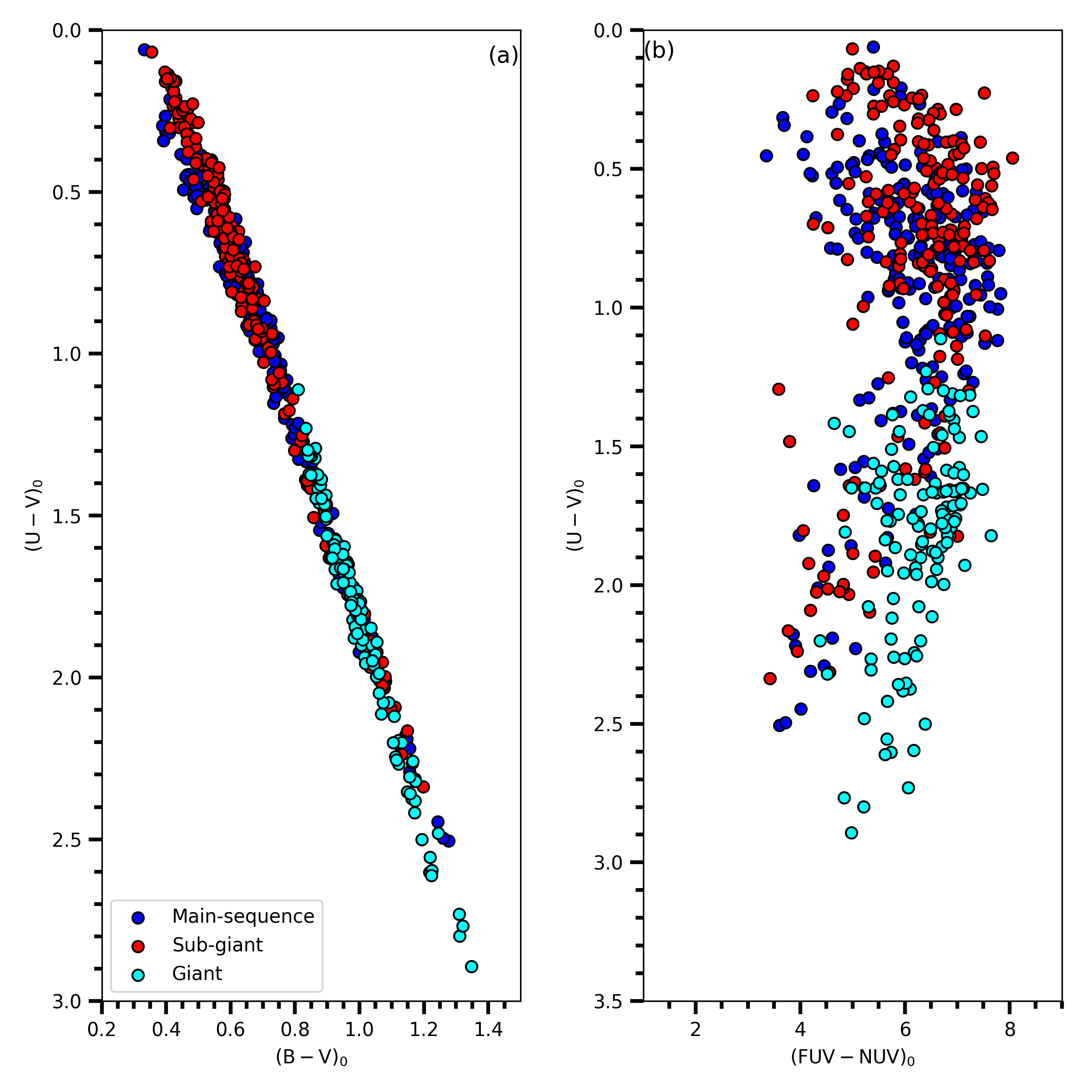}
\caption{Distribution of the sample stars in the $(U-V)_0\times(B-V)_0$ (a) and $(U-V)_0\times(FUV-NUV)_0$ (b) two-colour diagrams, colour coded for the luminosity class as indicated.}
\end{figure*} 

\begin{figure*}
\centering
\includegraphics[width=12cm,height=12cm,]{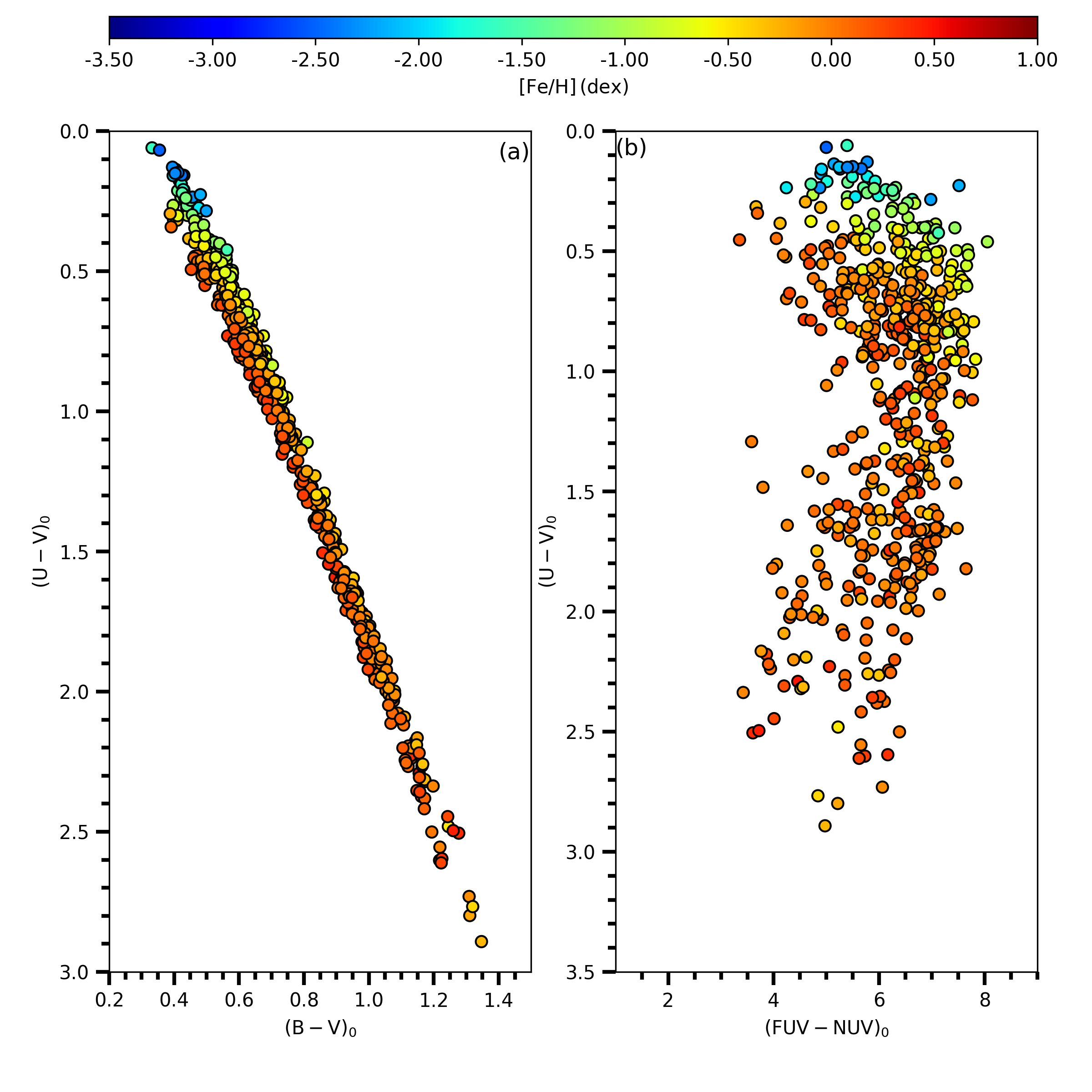}
\caption{Distribution of the sample stars in the $(U-V)_0\times(B-V)_0$ (a) and $(U-V)_0\times(FUV-NUV)_0$ (b) two-colour diagrams, colour coded for the metallicity as indicated.}
\end{figure*} 

\section{Transformation Equations}
We derived transformation equations between the $(FUV-NUV)_0$ and $(U-V)_0$ colour indices as a function of luminosity class as well as metallicity.

\subsection{Transformation Equations According to Luminosity Classes of Stars}
We adopted the following equation to transform the $(FUV-NUV)_0$ colour index to the $(U-V)_0$ as a function of the luminosity class,

\begin{equation}
(U-V)_0=a(FUV-NUV)_0+b(B-V)_0+c  
\end{equation} 

The numerical values of the coefficients $a$, $b$ and $c$ estimated for 245 main-sequence stars, 187 sub-giants and 124 giants by using multiple regression method are given in Table 4. $T$ and $p$ values corresponding to the sensitivity of the coefficients are given in the third and fourth lines for each coefficient, while the squared correlation coefficients and the standard deviations are given in the last two columns of the table. One can see that the correlation coefficients are rather high, while the standard deviations are small. The residuals for the colour index $(U-V)_0$, the differences between the original colour indices and the estimated ones are rather small and no systematic differences can be seen in their distribution (Fig. 8). However, there is an exception for the giants, i.e. the uncertainty for the coefficient a is larger than itself, additionally the corresponding $p$ value is greater than the usual one, $p=0.05$.

\begin{table*}[htbp]
  \centering
  \caption{Coefficients derived from Eq. (4) and the corresponding squared correlation coefficient ($R^2$) and standard deviation ($\sigma$), 
for the sample stars of different luminosity classes. The metallicities are not considered in these calculations. $N$ indicates the number of stars. The remaining symbols are explained in the text.}
    \begin{tabular}{ccccccc}
\hline
    Luminosity Class & $N$ & $a$ & $b$ & $c$ & $R^2$ & $\sigma$   \\
\hline
              &     & -0.042159(0.003659) & 2.65427(0.02125) & -0.63791(0.02827) &  &  \\
Main-sequence & 245 & $T$ = -11.52  & $T$ = 124.90 & $T$ = -22.57 & 0.985 & 0.055 \\
              &     & $p$ = 0.000   & $p$ = 0.000  & $p$ = 0.000  &  &  \\
\hline
              &  & -0.020667(0.004340) & 2.79537(0.02161) & -0.88118(0.03417) &  &  \\
Sub-giant     &  187   & $T$ = -4.76 & $T$ = 129.35  & $T$ = -25.79 & 0.990 & 0.055 \\
              &   & $p$ = 0.000 & $p$ = 0.000 & $p$ = 0.000 &   &  \\
\hline

             &     & 0.002630(0.007415) & 3.18474(0.04374) & -1.37664(0.07691) &  &  \\
 Giant       & 124 & $T$ = 0.35 & $T$ = 72.82 & $T$ = -17.90 & 0.982 & 0.050 \\
             &     & $p$ = 0.723  & $p$ = 0.000 & $p$ = 0.000 &   &  \\
\hline
    \end{tabular}%
  \label{tab:addlabel}%
\end{table*}%

\begin{figure*}[t]
\centering
\includegraphics[width=15cm,height=9cm,]{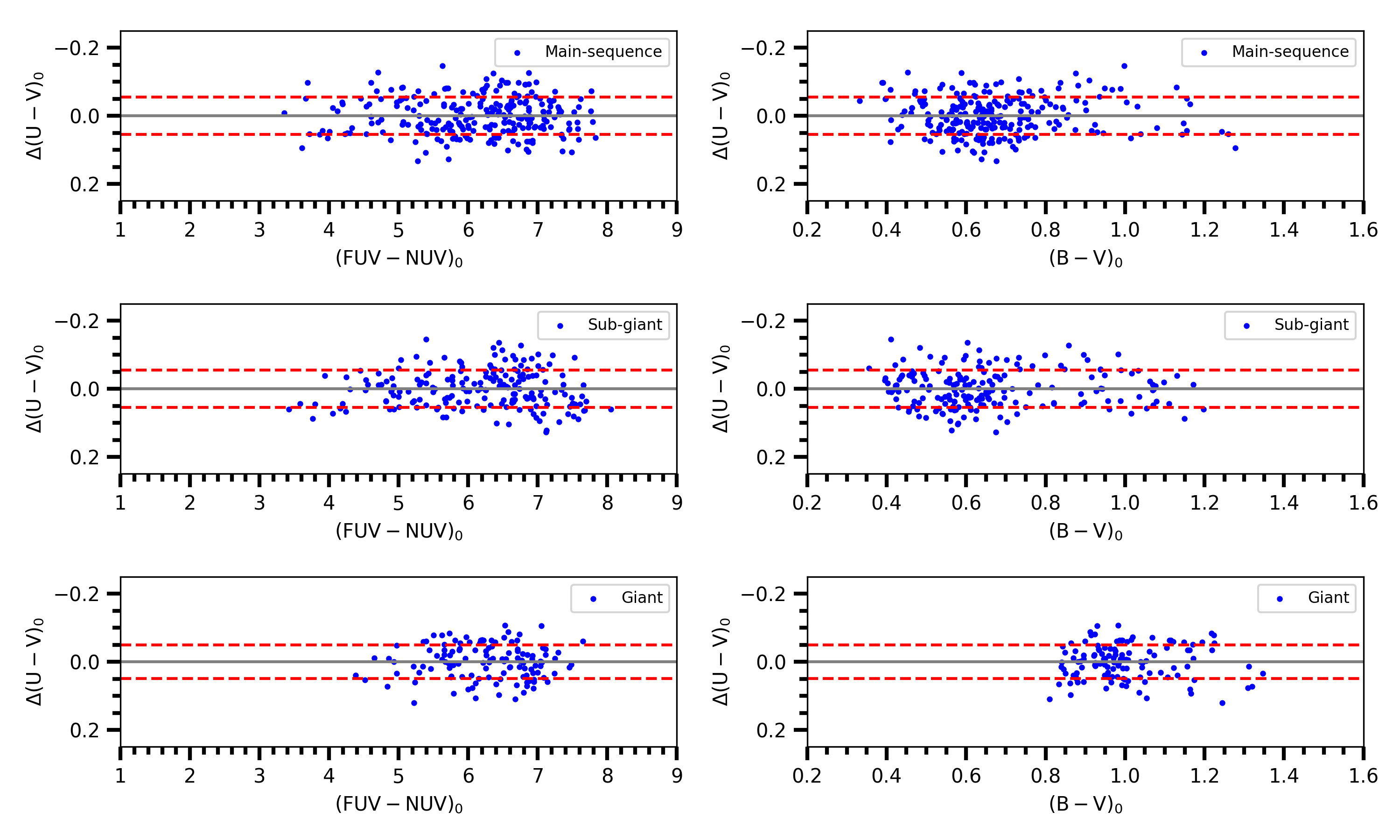}
\caption{Colour residuals in terms of $(FUV-NUV)_0$ (left column) and $(B-V)_0$ (right column) for three luminosity classes as indicated in six panels. Metallicity is not considered in calculation of the residuals. Dashed lines denote $\pm 1\sigma$ prediction levels.}
\end{figure*} 

\begin{table*}[htbp]
{\small
  \centering
  \caption{Coefficients derived from Eq. (4) and the corresponding statistical results for sample stars of different luminosity classes and metallicities. $N$ indicates the number of stars. The remaining symbols are explained in the text.}
    \begin{tabular}{cccccccc}
\hline
Luminosity Class & [Fe/H]    & $N$ & $a$ & $b$ & $c$ & $R^2$ & $\sigma$ \\
                 & (dex)     &     &     &     &     &       &    \\
\hline
                 & (0, +0.5] & 110 & -0.021402(0.004368) & 2.62304(0.02228)  & -0.70015(0.03360) & 0.993 & 0.043\\
                 &           &     & $T$= -4.90          & $T$= 117.72       & $T$= -20.84       &       &  \\
                 &           &     & $p$=  0.000         & $p$= 0.000        & $p$= 0.000        &       &  \\

Main-sequence    & (-0.5, 0] & 102 & -0.034443(0.004107) & 2.58820(0.02807)  & -0.66283(0.03011) & 0.989 & 0.038 \\      
                 &           &     & $T$ = -8.39         & $T$ = 92.22       & $T$ = -22.01      &       &  \\
                 &           &     & $p$ = 0.000         & $p$ = 0.000       & $p$ = 0.000       &       &  \\

                 & (-1, -0.5]& 26  & -0.03266(0.01207)   & 2.4366(0.1004)    & -0.63172(0.05707) & 0.979 & 0.031 \\
                 &           &     & $T$ = -2.71         & $T$ = 24.27       & $T$ = -11.07      &       &  \\
                 &           &     & $p$ = 0.013         & $p$ = 0.000       & $p$ = 0.000       &       &  \\
\hline
\hline
                 & (0, +0.5] & 50 & -0.002031(0.000630)  & 2.76866(0.03730) & -0.92365(0.05313)  & 0.992 & 0.040 \\
                 &           &    & $T$ = -2.48          & $T$ = 74.22      & $T$ = -17.38       &       &  \\
                 &           &    & $p$ = 0.014          & $p$ = 0.000      & $p$ = 0.000        &       &  \\
                 & (-0.5, 0] & 64 & -0.012175(0.005475)    & 2.77782(0.02874) & -0.93117(0.04992)& 0.996 & 0.036 \\
                 &           &    & $T$ = -2.22          & $T$ = 96.65      & $T$ = -18.65       &       &  \\
                 &           &    & $p$ = 0.030          & $p$ = 0.000      & $p$ = 0.000        &       &  \\
Sub-giant        & (-1, -0.5]& 35 & -0.014890(0.008362)  & 2.04629(0.08635) & -0.52484(0.04988)  & 0.959 & 0.029 \\
                 &           &    & $T$ = -2.15          & $T$ = 23.70      & $T$ = -10.52       &       &  \\
                 &           &    & $p$ = 0.039          & $p$ = 0.000      & $p$ = 0.000        &       &  \\
                 & (-3, -1]  & 38 & -0.004800(0.000810)  & 1.9139(0.1289)   & -0.59024(0.04176)  & 0.926 & 0.030 \\
                 &           &    & $T$ = -1.59          & $T$ = 14.85      & $T$ = -14.13       &       &  \\
                 &           &    & $p$ = 0.038          & $p$ = 0.000      & $p$ = 0.000        &       &  \\
\hline
\hline
                 & (0, +0.5] & 57 & 0.003814(0.007765)   & 3.23331(0.04865) & -1.39634(0.07903)  &0.989  & 0.034 \\
                 &           &    & $T$ = 0.49           & $T$ = 66.47      & $T$ = -17.67       &       &  \\
 Giant           &           &    & $p$ = 0.625          & $p$ = 0.000      & $p$ = 0.000        &       &  \\
                 & (-0.5, 0] & 62 & -0.004697(0.006729)  &3.07446(0.04036)  & -1.24802(0.07181)  & 0.993 & 0.034 \\
                 &           &    & $T$ = -0.70          & $T$ = 76.17      & $T$ = -17.38       &       &  \\
                 &           &    & $p$ = 0.488          & $p$ = 0.000      & $p$ = 0.000        &       &  \\
\hline

    \end{tabular}%
  \label{tab:addlabel}%
}
\end{table*}%

\begin{landscape}
\textwidth = 750pt
\begin{figure*}
\includegraphics[width=26cm,height=14cm,keepaspectratio]{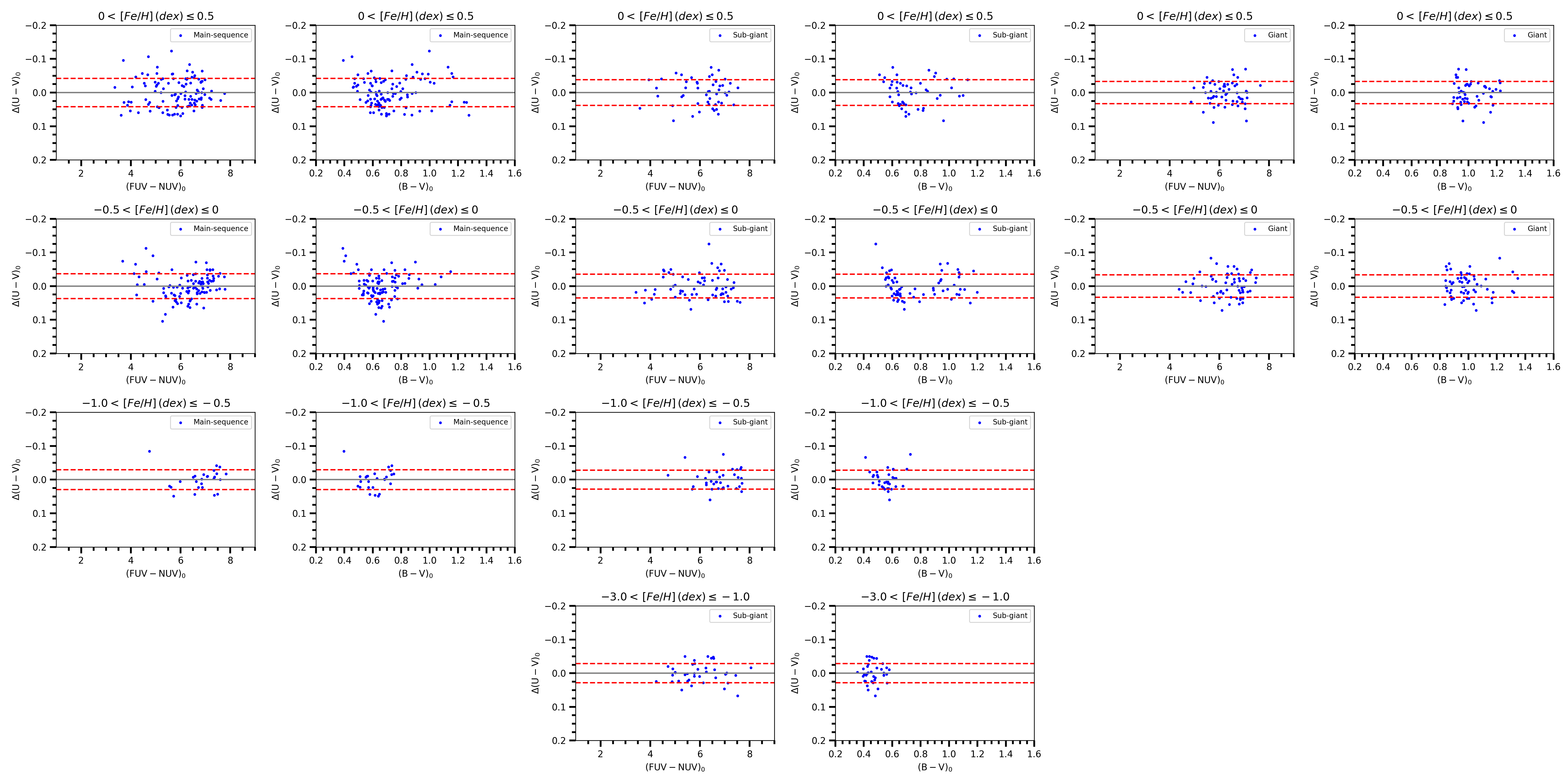}
\caption{Colour residuals for the sample stars in terms of $(FUV-NUV)_0$ and $(B-V)_0$ colours for different luminosity classes and metallicities, as indicated in the panels. Dashed lines denote $\pm 1\sigma$ prediction levels.}
\end{figure*} 
\end{landscape}

\subsection{Transformation Equations According to the Luminosity Classes and Metallicities of Stars} 
The sample stars are separated into different metallicity intervals and transformation equations are derived for three luminosity classes of stars in each metallicity interval, as explained in the following. Main-sequence and sub-giant stars occupy the metallicity intervals $0<{\rm [Fe/H]}\leq+0.5$ dex, $-0.5<{\rm [Fe/H]}\leq0$ dex, $-1<{\rm [Fe/H]}\leq-0.5$ dex, additionally sub-giants occupy the interval $-3<{\rm [Fe/H]}\leq-1$ dex, while giants cover the metallicity intervals $0<{\rm [Fe/H]}\leq0.5$ dex, and $-0.5<{\rm [Fe/H]}\leq0$ dex. We used the Eq. (4) and estimated the numerical values of the coefficients $a$, $b$, and $c$ for each metallicity interval and luminosity class by the procedure explained in Section 3.1. Result are tabulated in Table 5. The squared correlation coefficients in this table are higher than those in Table 4. Also, the standard deviations in Table 4 reduced by 30\%. Comparison of the observed $(U-V)_0$ colour indices and the estimated ones via Eq. (4) are given in Fig. 9. As it can be seen easily, there is no any systematic deviations in the distribution of residuals for $(U-V)_0$ colour index. They are smaller than the those plotted in Fig. 8, as well. However, we should note that the coefficient $a$ estimated for the metallicity intervals of giants does not promise accurate $(U-V)_0$ colour index estimation. 

\section{Summary and Discussion}
In this study, we used 556 stars with accurate spectroscopic, photometric and astrometric data and derived transformation equations between {\it GALEX} and {\it UBV} colours. Thus, the $U$ magnitudes of the stars would be estimated more accurately by means of $FUV$ and $NUV$ magnitudes which are observed outside of the Earth atmosphere. Transformation equations are derived as a function of (only) the luminosity class, and as a function of both the luminosity class and metallicity. In both cases the statistical results promise accurate $U-V$ colours for the main-sequence and sub-giant stars, estimated by using the $FUV$ and $NUV$ magnitudes. However, the same case does not hold for the giants.

We used the inverted parallaxes as a distance estimate when calculating the interstellar absorption. However, \citet*{Schonrich19} have shown that the {\it Gaia} DR2 parallaxes can be biased. We compared the distances estimated via {\it Gaia} DR2 trigonometric parallaxes and the ones of \citet{Schonrich19} which are obtained by a statistical method to see the impact of the distances of the stars on the analysis, as explained in the following. The mean of the differences between the distances estimated by two procedures and the corresponding standard deviation are -0.10 pc and 0.66 pc, respectively. As seen Fig. 10, almost all distances fit with the one-to-one straight line. Hence, we do not expect any systematic uncertainty for our results.

\begin{figure*}
\centering
\includegraphics[width=10cm,height=10cm,keepaspectratio]{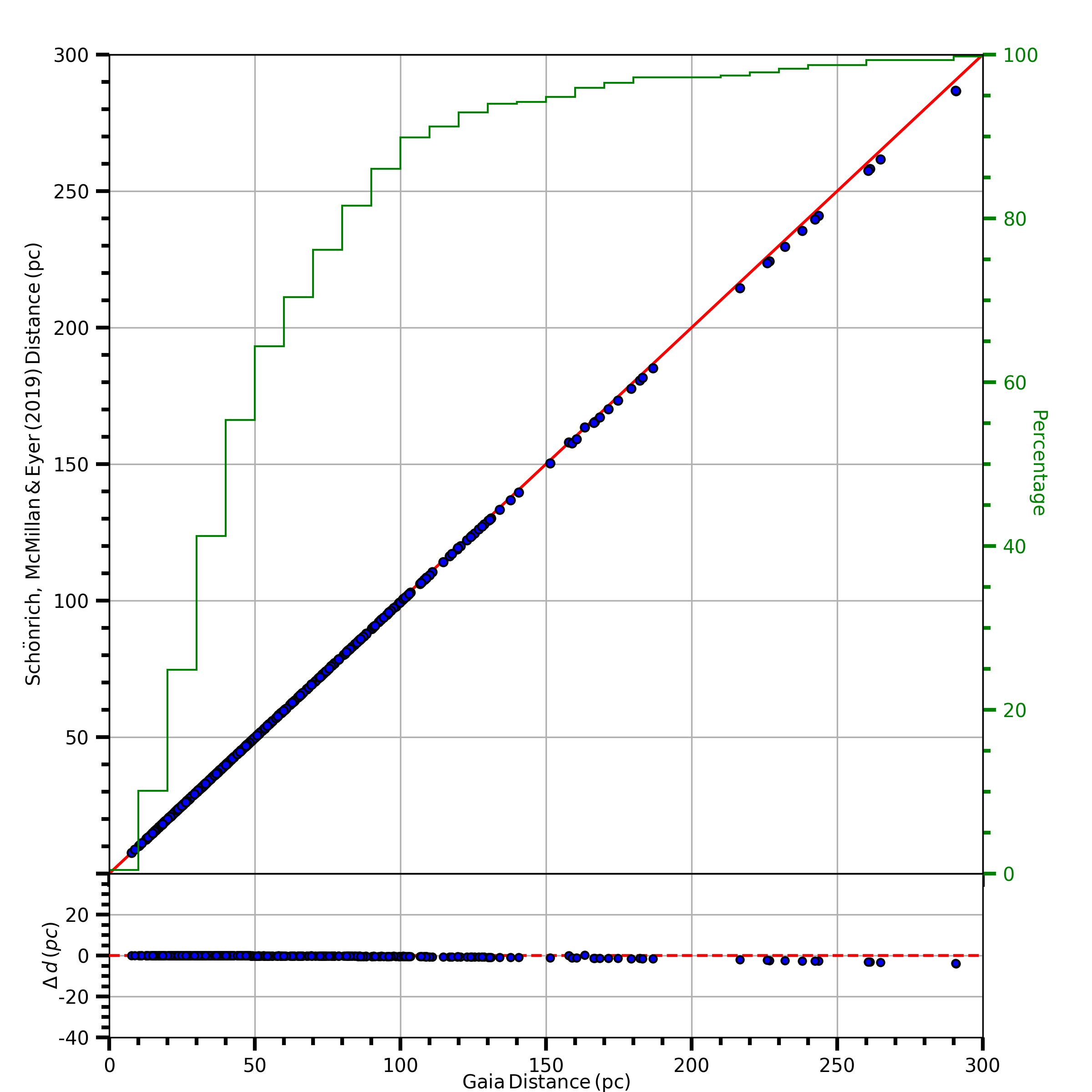}
\caption{Comparison of the distances for the sample stars estimated via {\it Gaia} DR2 trigonometric parallaxes and statistical method of \citet{Schonrich19}. The distances calculated by two different methods are quite compatible with one-to-one line.}
\end{figure*} 

We compared the spectroscopic atmospheric model parameters taken from 14 different papers in the literature to investigate the confirmation of their homogeneity. Figs. A1, A2 and A3 show that the three parameters, $T_{eff}$, $\log g$, and [Fe/H] lie on the one-to-one line in the corresponding figure. Hence, we can argue that the atmospheric model parameters taken from different studies are in an agreeable homogeneity.

The transformation equations are applied to the F-G type main-sequence stars \citet{Tuncel16} which are provided with accurate photometric data. The sample of stars in the cited study reduced from 168 to 70 due to the absence of $FUV$ and $NUV$ magnitudes in {\it GALEX} DR7 \citep{Bianchi17} database. We used the corresponding coefficients in Table 4 and Table 5, and estimated the $(U-V)_0$ colours of 70 stars in question. Residuals (Fig. 11) and the statistical results (Table 6) show that combination of the luminosity class and metallicity provides more accurate $(U-V)_0$ colours relative to the ones estimated by considering only the luminosity class. We should emphasize that the results corresponding only to the  luminosity class are also consistent.
 
\begin{table*}[htbp]
  \centering
\setlength{\tabcolsep}{3pt}
{\scriptsize
  \caption{Statistical results based on the comparison of the observed and calculated $(U-V)_0$ colours according to the coefficients in Tables 4 and 5 (sum of differences ($\Sigma(\Delta (U-V)_0)$), means of differences ($\Sigma(\Delta (U-V)_0)/N$), and standard deviations of differences $\sigma_{\Sigma(\Delta (U-V)_0)/N}$) for 70 main-sequence stars.}
    \begin{tabular}{cccccccc}
\hline
      & & \multicolumn{3}{c}{Coefficients in Table 4} & \multicolumn{3}{c}{Coefficients in Table 5}\\
\hline
${\rm [Fe/H]}$ (dex) & $N$ &  $\Sigma(\Delta (U-V)_0)$ & $\Sigma(\Delta (U-V)_0)/N$ & $\sigma_{\Sigma(\Delta (U-V)_0)/N}$ &  $\Sigma(\Delta (U-V)_0)$ & $\Sigma(\Delta (U-V)_0)/N$ & $\sigma_{\Sigma(\Delta (U-V)_0)/N}$  \\
\hline 
$0<{\rm [Fe/H]}\leq0.5$   &  7 &  0.204 &  0.029 & 0.035 &-0.159 &-0.023 & 0.031 \\
$-0.5<{\rm [Fe/H]}\leq0$  & 38 & -0.501 & -0.013 & 0.044 & 0.059 & 0.002 & 0.041 \\
$-1<{\rm [Fe/H]}\leq-0.5$ & 25 & -0.837 & -0.033 & 0.051 & 0.422 & 0.017 & 0.042 \\
\hline
    \end{tabular}%
  \label{tab:addlabel}%
}
\end{table*}%

The transformation equations between the {\it GALEX} and {\it UBV} colours would be used for estimation of the $U$ magnitude of stars for which this magnitude cannot be observed accurately. This is important for the intermediate spectral type main-sequence stars. Because, the $U$ magnitude thus obtained, i.e. $U_{est}$, would be combined with the $B$ magnitude of {\it UBV} photometric system and the $U_{est}-B$ colour would be used in the (photometric) metallicity estimation which is important in studying the chemical structure and evolution of our Galaxy.

\begin{figure*}
\centering
\includegraphics[width=14.5cm,height=14.5cm,keepaspectratio]{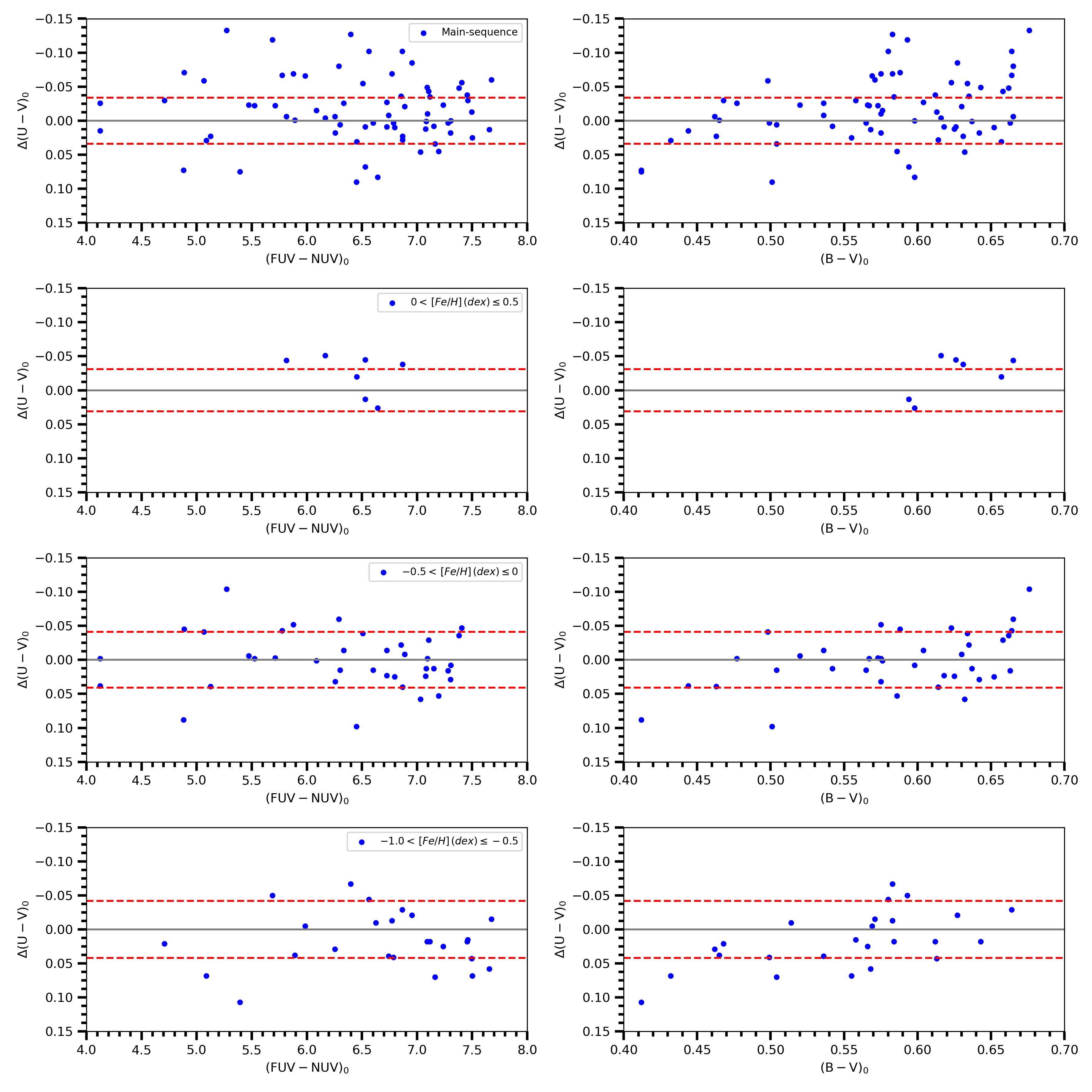}
\caption{Colour residuals for 70 main-sequence stars taken from \citet{Tuncel16} in terms of $(FUV-NUV)_0$ (left column) and $(B-V)_0$ (right column). Residuals for stars with different metallicities are indicated in the panels. Dashed lines show $\pm 1\sigma$ prediction levels.}
\end{figure*} 

\section{Acknowledgments}
The authors thank the anonymous referee for her/his suggestions that helped improve the quality of the paper. This research has made use of NASA's (National Aeronautics and Space Administration) Astrophysics Data System and the SIMBAD Astronomical Database, operated at CDS, Strasbourg, France and NASA/IPAC Infrared Science Archive, which is operated by the Jet Propulsion Laboratory, California Institute of Technology, under contract with the National Aeronautics and Space Administration. This work has made use of data from  the European Space Agency (ESA) mission {\it Gaia} (\mbox{https://www.cosmos.esa.int/gaia}), processed by the {\it Gaia} Data Processing and Analysis Consortium (DPAC, \mbox{https://www.cosmos.esa.int/web/gaia/dpac/consortium}). Funding for the DPAC has been provided by national institutions, in particular the institutions participating in the {\it Gaia} Multilateral Agreement.

\begin{appendix}

\begin{figure*}
\centering
\includegraphics[width=15cm,height=15cm,keepaspectratio]{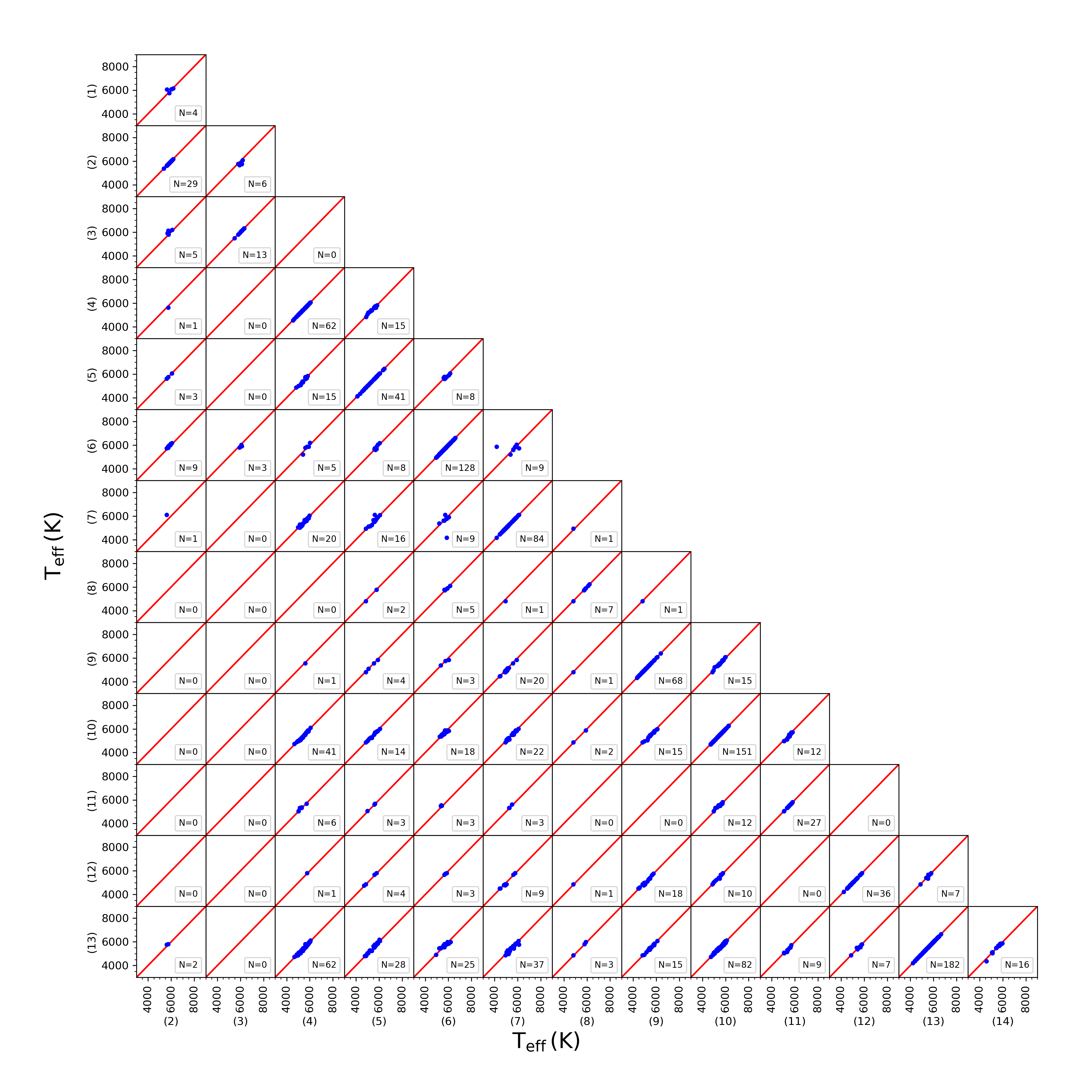}
\caption{Corner plot showing a comparison of $T_{eff}$ for overlapping 
stars in 14 research groups. The numbers indicate the research  groups: 
(1) \citet{Boesgaard11}, (2) \citet{Nissen11}, (3) \citet{Ishigaki12}, 
(4) \citet{Mishenina13}, (5) \citet{Molenda13}, (6) \citet{Bensby14}, 
(7) \citet{daSilva15}, (8) \citet{Sitnova15}, (9) \citet{Jofre15}, 
(10) \citet{Brewer16}, (11) \citet{Kim16}, (12) \citet{Maldonado16}, 
(13) \citet{Luck17}, (14) \citet{DelgadoMena17}.} 
\end{figure*}

\begin{figure*}
\centering
\includegraphics[width=15cm,height=15cm,keepaspectratio]{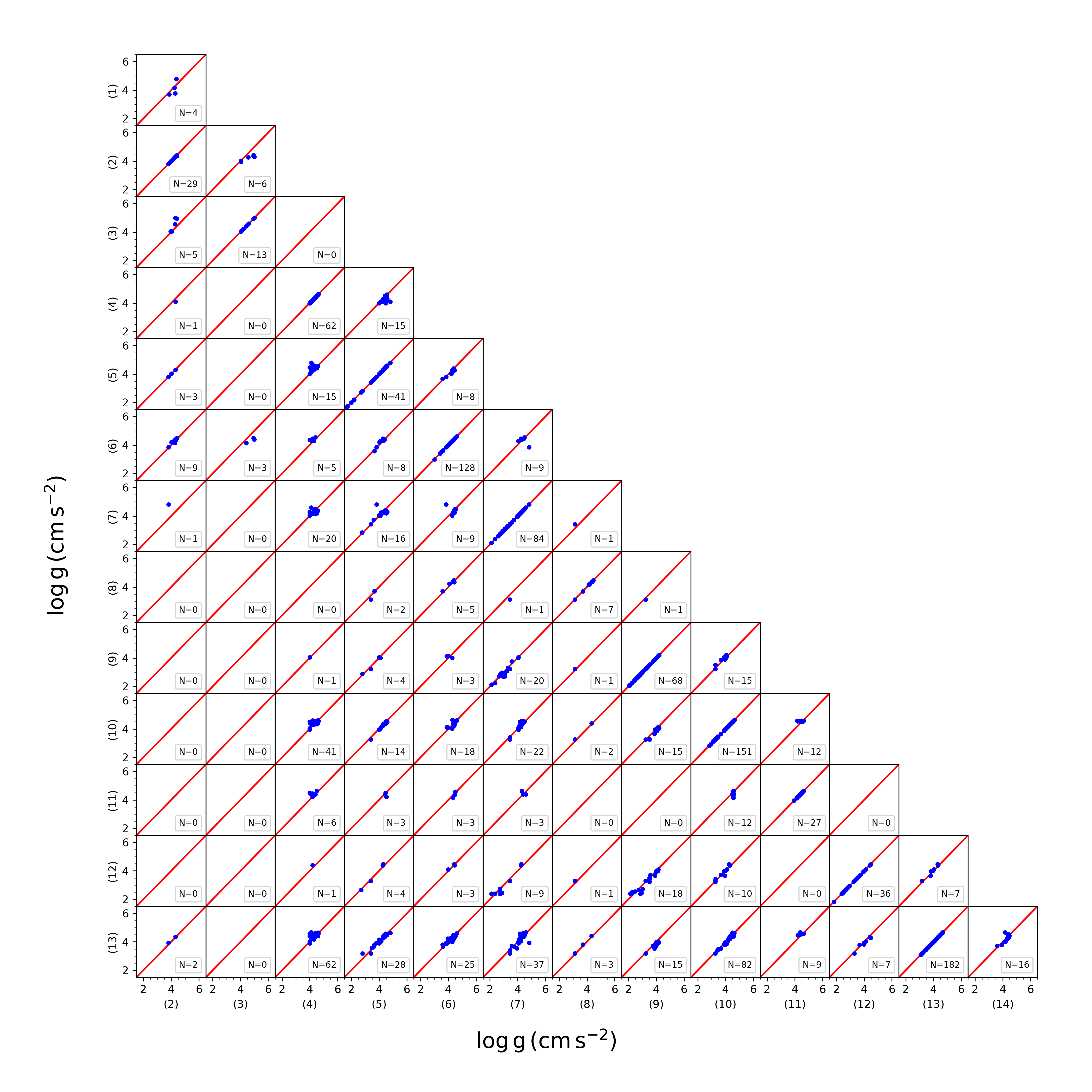}
\caption{Corner plot showing a comparison of $\log g$ for overlapping 
stars in 14 research groups. The numbers indicate the the research 
groups as in Fig. A1.} 
\end{figure*}

\begin{figure*}
\centering
\includegraphics[width=15cm,height=15cm,keepaspectratio]{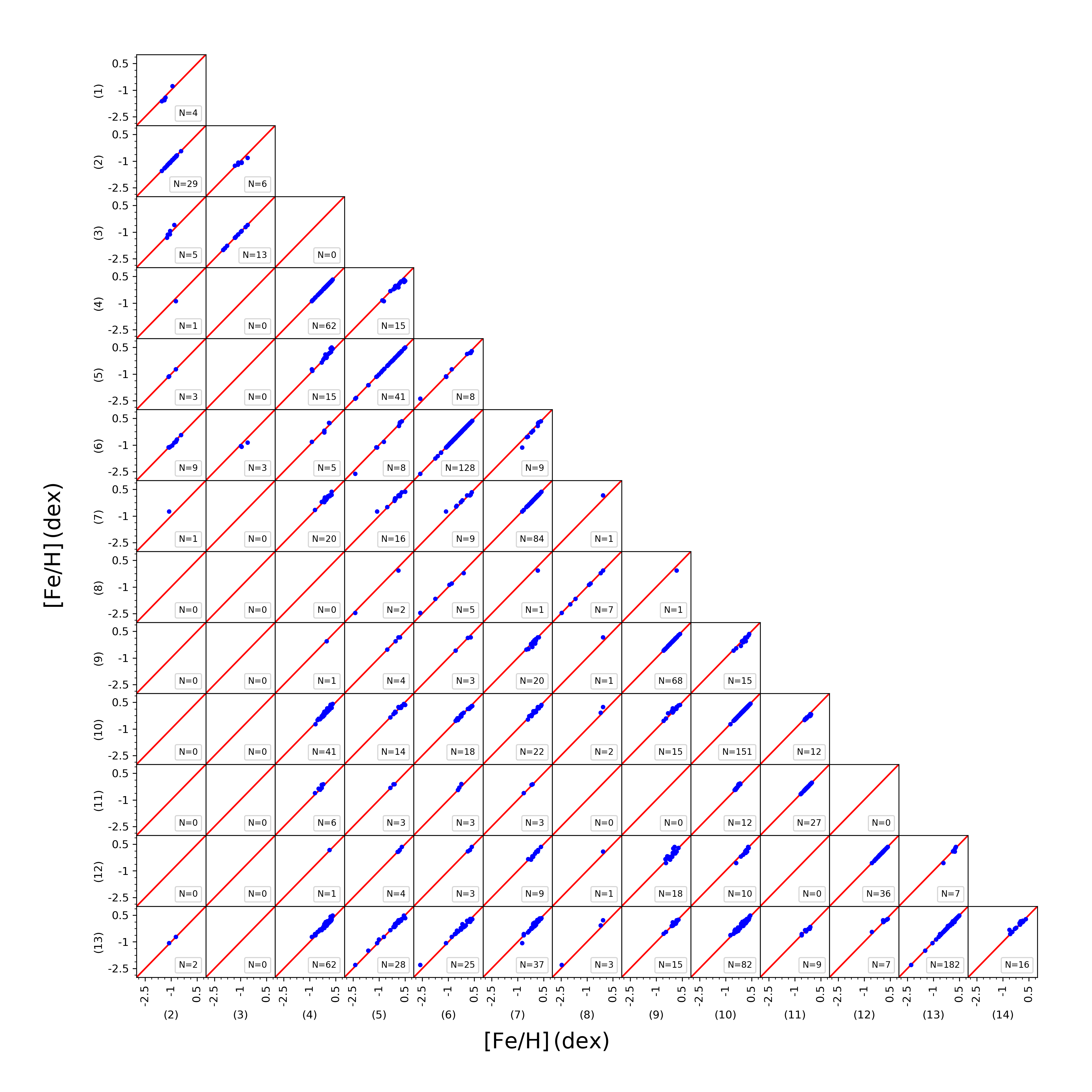}
\caption{Corner plot showing a comparison of [Fe/H] for overlapping stars 
in 14 research groups. The numbers indicate the research groups as in 
Fig. A1.} 
\end{figure*}

\end{appendix}

\end{document}